\documentclass[aps,pra,twocolumn,showpacs,preprintnumbers,amsmath,amssymb,longbibliography]{revtex4-2}
\usepackage{graphicx}
\usepackage[usenames,dvipsnames]{xcolor}
\usepackage{amsmath}
\usepackage{cleveref}
\usepackage{makecell}
\usepackage{float}
\usepackage{lineno}
\usepackage[caption=false]{subfig}
\newcommand{\Tr}{\textrm{Tr}}

\newcommand{\bra}[1]{\ensuremath{\langle #1 |}}
\newcommand{\ket}[1]{\ensuremath{| #1 \rangle}}

\renewcommand{\dag}{\dagger}
\usepackage{xcolor}

\usepackage{ulem}
\usepackage{amsmath,amssymb}
\begin{document}

\title{Calculating the Luttinger liquid parameter for an interacting Kitaev chain \\ quantum simulator}

\author{Troy Losey$^1$}
\email{tlose001@ucr.edu}
\author{Jin Zhang$^2$}
\email{jzhang91@cqu.edu.cn}
\author{S.-W. Tsai$^1$}
\affiliation{$^1$ Department of Physics and Astronomy, University of California, Riverside, California 92521, USA}
\affiliation{$^2$ Department of Physics and Chongqing Key Laboratory for Strongly Coupled Physics, Chongqing University, Chongqing 401331, China}
\definecolor{burnt}{cmyk}{0.2,0.8,1,0}
\def\lt{\lambda ^t}
\def\note{note}
\def\beq{\begin{equation}}
\def\enq{\end{equation}}

\date{\today}
\begin{abstract}

In this work, we introduce a solid-state platform for building quantum simulators using implanted spin centers in solid-state materials. We build upon the proposal for an $S=1$ chain of spin centers coupled through the magnetic dipole-dipole interaction and subjected to an external magnetic field as a quantum simulator for critical floating phases. We introduce another magnetic field and map the system to the interacting Kitaev chain. This setup, tunable through the applied fields and the orientation of the spin centers within the crystal, exhibits a variety of rich quantum behavior which notably includes floating phases, a $Z_2$ symmetry-breaking phase, and lines of both Berezinskii–Kosterlitz–Thouless (BKT) and Pokrovsky–Talapov transitions. Furthermore, we employ several novel methods to calculate the Luttinger liquid parameter in our model with incommensurate correlations. We find that these methods provide a route to identify BKT transitions with less computational resources than utilizing entanglement entropy and central charge.

\end{abstract}

\maketitle

\section{Introduction}
\label{sec:introduction3}

Quantum simulators have long been proposed as powerful tools to study quantum models that are too complex for classical computers to simulate~\cite{feynman2018simulating}. Analog quantum simulators utilizing quantum dots~\cite{barthelemy2013quantum} and ultracold atoms, like Rydberg atoms~\cite{browaeys2020many}, have already demonstrated the ability to realize interesting quantum models~\cite{hensgens2017quantum,bernien2017probing,barredo2018synthetic}. Recently, spin centers implanted in solid-state hosts have emerged as a promising alternative for quantum simulation~\cite{losey2024quantum}. Spin centers are defects in crystals that behave as localized spin-S particles, which can be optically initialized with lasers and read out via photoluminescence with high fidelity. Spin centers show great promise for both quantum information and quantum sensing applications because they have been found to have long coherence times even at room temperature, and their energy levels are especially sensitive to electric and magnetic fields~\cite{weber2010quantum,schirhagl2014nitrogen}. Furthermore, the spin-spin interactions between spin centers naturally support the creation of spin array Hamiltonians that can give rise to many critical and exotic behaviors. Consequently, spin centers offer a robust and scalable route to explore spin Hamiltonians whose emergent phenomena embody universal features of quantum criticality.

Despite these advantages, engineering strong and controllable interactions between spin centers remains a major challenge. The same electronic isolation that preserves coherence limits dipolar and exchange couplings to relatively weak and short-range interactions. However, recent proposals to mediate spin-spin couplings through magnonic modes~\cite{candido2020predicted,fukami2021} suggest a path forward, where engineered effective couplings between spin centers enable interactions that are stronger at up to micrometer distances than what is currently achievable with $10$~nm spin center separations. Consequently, there is potential for spin centers implanted in a 3D crystal to become powerful tools to simulate exotic quantum phases and phase transitions in up to three dimensions, as well as feasible candidates for room-temperature quantum computing. Furthermore, there are a plethora of spin center candidates with varying electronic properties that can be utilized to simulate spin-S spin models and perform quantum computations as qudits~\cite{gali2019ab,tarasenko2018spin,azamat2012electron}. Additionally, there have been proposals to realize non-abelian anyons with quantum dot \cite{cookmeyer2024engineering}, ultracold atom \cite{duan2003controlling}, Floquet drive \cite{sun2023engineering}, and trapped ion \cite{schmied2011quantum} systems, and spin center-based quantum simulators may provide another route to realizing non-abelian anyons. Meaningful proposals to employ spin centers for quantum information applications are necessary to push the development of experimental techniques and engineered interactions between spin centers to meet these long-term goals for quantum information.

In this paper, we propose a new magnetic field configuration for the magnetic dipole-dipole coupled effective spin-1/2 spin center chain quantum simulator that was recently proposed as the first solid-state quantum simulator for critical floating phases~\cite{losey2024quantum}. This new simulator is described by an XYZ Heisenberg spin chain with external magnetic fields, where we control the fields to impose a $Z_2$ symmetry, and accordingly, the model maps exactly to the interacting Kitaev chain. We reveal that this new magnetic field configuration allows for the quantum simulation of a $Z_2$ symmetry-breaking phase and provides a substantial improvement to the quantum simulator's ability to realize floating phases and both Berezinskii–Kosterlitz–Thouless (BKT) and Pokrovsky–Talapov (PT) transitions. The floating phases are exotic gapless magnetic phases characterized by incommensurate quasi-long-range order. In 1-dimensional systems, floating phases are described by the Tomonaga–Luttinger liquid (TLL) model, which describes the low-energy properties of many gapless 1-dimensional systems. The TLL model becomes unstable at a BKT transition, which is an infinite-order phase transition between Luttinger liquid and bulk disordered phases. Notably, the $Z_2$ symmetry-breaking phase is a symmetry-protected topological (SPT) phase with gapless edge modes when the interaction coefficient $U$ [see Eq.~\ref{eq:KitaevChain}] goes to zero and the Kitaev chain becomes non-interacting. With our proposed setup, $U/t \geq 1/2$; however, future spin center-based quantum simulators may be able to realize this SPT phase.

Central charge and the scaling of entanglement entropy as the system size increases usually provide reliable ways to identify critical points and phases; however, this is not generally true within floating phases. An incompatibility between the incommensurate order and finite system size causes discrete changes in the wave vector of the floating phase ground states as the system size changes. This makes it challenging to view the scaling of entanglement entropy or precisely calculate the central charge within this phase. We show that if the magnetization of the ground state is held constant, the entanglement entropy increases smoothly with system size, and the central charge can be extracted with great precision. Nevertheless, we find that at finite system sizes, the central charge $c=1$ within both the floating phases and $Z_2$ symmetry-breaking phases; therefore, entanglement entropy and central charge are still not reliable indicators for the BKT transitions in this model. However, since BKT transitions occur when the Luttinger liquid parameter $K$ reaches a critical value that destabilizes the TLL model, $K$ can be an effective way to identify BKT transitions. Consequently, we employ methods from conformal field theory (CFT) to utilize Friedel oscillations~\cite{white2002friedel} and crosscap states~\cite{tan2025extracting} to extract the Luttinger liquid parameter $K$ from ground state wave functions. We find that extracting $K$ from the crosscap overlap can be challenging in the floating phases, where contributions to the overlap from the ground state's different particle number sectors can interfere with each other. Nevertheless, we resolve this complication by calculating and utilizing the energy spectrum at the PT transition to choose ground states with large weights for the desired particle number sectors.

The paper is organized as follows. In Section~\ref{sec:model3}, we summarize the derivation of the Hamiltonian for the magnetic dipole-dipole coupled effective spin-1/2 spin center chain and provide a mapping to the interacting Kitaev model. Furthermore, we highlight notable experimental considerations for realizing spin center arrays. A more detailed derivation of the Hamiltonian and discussion of experimental feasibility are provided in Ref.~\cite{losey2024quantum}. In Section~\ref{sec:results3}, we present a phase diagram for this model and show how both PT and BKT lines are found. Furthermore, we compare several methods to extract the Luttinger liquid parameter $K$ and derive an expression for the energy spectrum at a PT transition to better calculate $K$. We summarize the results of the paper and discuss future outlooks in Sec.~\ref{sec:conclusion3}.

\section{Model}
\label{sec:model3}

\begin{figure*}[t!]
  \begin{center}
    \includegraphics[width=\textwidth]{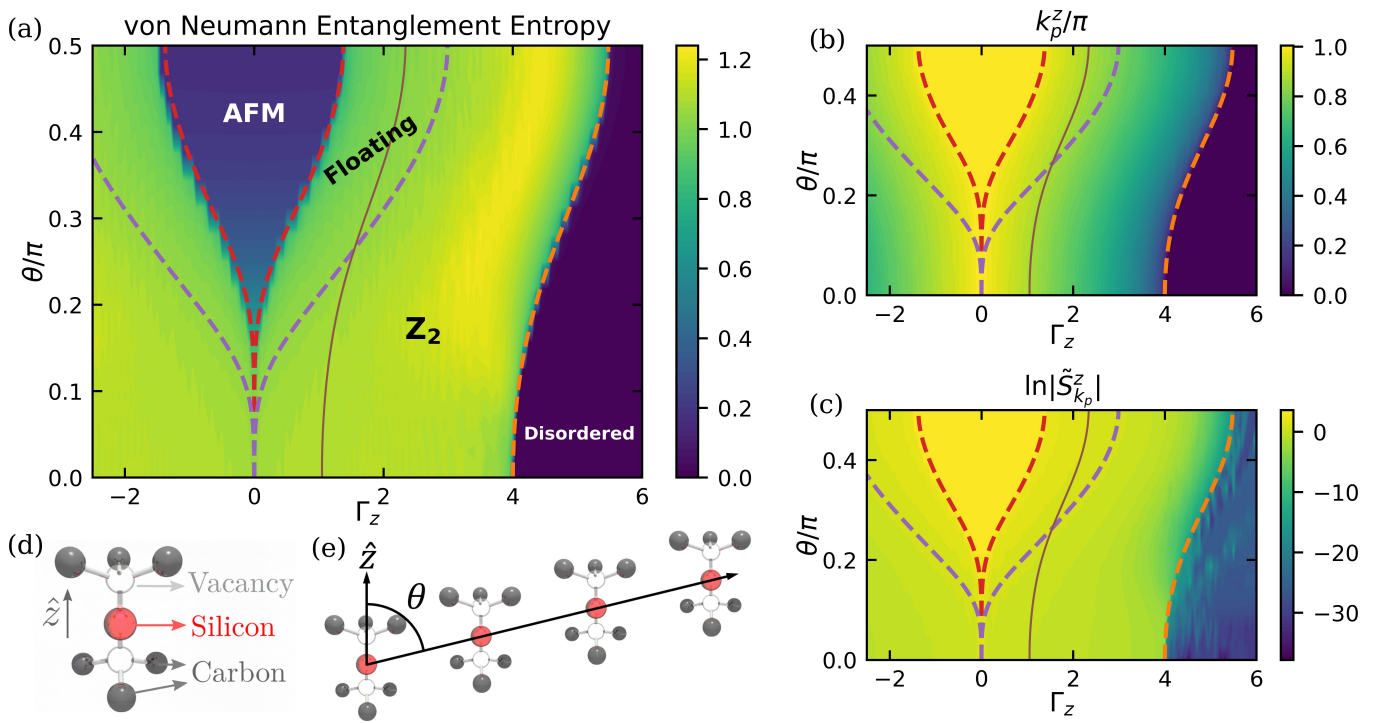}
    \caption{\label{fig:phaseDiagram}Subplot (a) depicts a phase diagram showing the von Neumann entanglement entropy for the dipolar-coupled spin center chain portrayed in subplots (d) and (e). The angle $\theta$ defines the orientation of the chain relative to the spin centers' $\hat{z}$ symmetry axes, and $\Gamma_z$ is an effective magnetic field. The low entanglement entropy region between the red dashed lines is an AFM phase, while the red dashed lines are PT transitions. The purple dashed lines show BKT transitions, and the green region between the PT and BKT lines are critical floating phases. The orange dashed line is known analytically as the disorder line, and the high entanglement entropy region between the BKT and disorder lines is a $Z_2$ symmetry-breaking phase. The low entanglement entropy region for $\Gamma_z$ greater than the disorder line is a disordered phase. Along the cut where $\theta=0$, the $SU(2)$ symmetric Heisenberg point is the BKT point at $\Gamma_z=0$, there is a PT point at $\Gamma_z=4$, and between these points is a pure Luttinger liquid phase. The solid brown line is where the magnetization $m^z=1/16$. The phase diagram is reflected over $\Gamma_z=0$, $\theta=0$, and $\theta=\pi/2$. Subplots (b) and (c) show order throughout the phase diagram by plotting the wave vector $k^z_p$ that describes oscillations in the correlations, and the Fourier transform $\widetilde{S}^z_{k_p}$ at wave vector $k^z_p$ of the local expectation values $\langle S^z_i\rangle$, respectively.}
  \end{center}
\end{figure*}

\begin{figure}[t]
\centering
\includegraphics[width=.48\textwidth]{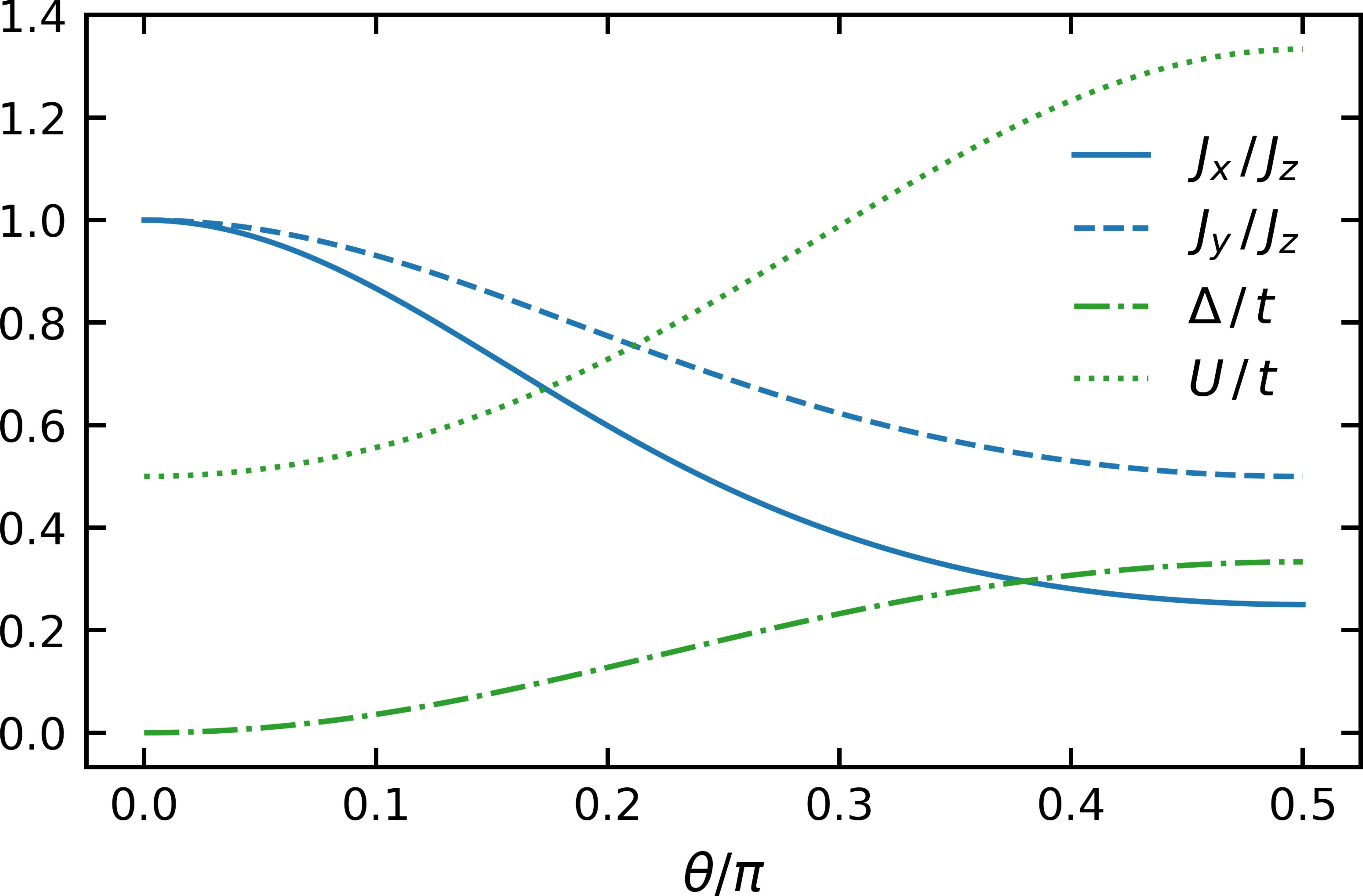}
\caption{\label{fig:HamCouplings}Hamiltonian parameters changing with the angle $\theta$ [see Fig.~\ref{fig:phaseDiagram}(e)]. The parameters are shown for the spin representation [see Eq.~(\ref{eq:rotatedHam3})] and the spinless fermionic representation [see Eq.~(\ref{eq:KitaevChain})].}
\end{figure}

In a previous work, we derived the mapping from a 1-dimensional chain of spin-1 spin centers with anisotropy that interact via the magnetic dipole-dipole interaction, to an effective spin-1/2 XYZ Heisenberg spin chain with an external field~\cite{losey2024quantum}. The mapping requires all spin centers to have their symmetry axes aligned with the same crystal axis. Moreover, the zero-field splitting Hamiltonian term must be much larger than the dipolar interaction term, which is true for most spin centers. These two assumptions allow an external magnetic field to make the $|0\rangle$ and $|-1\rangle$ spin states for each spin-1 spin center nearly degenerate, while the $|1\rangle$ spin state for each spin center is raised high enough in energy to be neglected, as the probability to enter the $|1\rangle$ spin state is suppressed at low temperatures. The effective spin-1/2 Hamiltonian is tunable via additional applied magnetic fields $h^z$ and $h^x$, and setting the angular displacement $\theta$ between the shared symmetry axes of the spin centers and the orientation of the spin chain [see Fig.~\ref{fig:phaseDiagram}(d,e)]. Here, we consider the additional applied magnetic fields to be small magnetic fields applied parallel ($h^z$) and perpendicular ($h^x$) to the shared spin center symmetry axes. Furthermore, $h^x$ is oriented such that it aligns with the spin center chain when $\theta=0$. The resulting Hamiltonian reads 
\begin{eqnarray}
\label{eq:rotatedHam3}
\nonumber H &=& \sum_{i=1}^{N-1} \left( J_x \sigma^x_i \sigma^x_{i+1} + J_z \sigma^z_{i} \sigma^z_{i+1} -\sigma^y_i \sigma^y_{i+1}\right) \\ &-& \sum_{i=1}^{N} \left( \Gamma_z\sigma^z_i + \Gamma_x\sigma^x_i \right),
\end{eqnarray}
with parameters
\begin{eqnarray}
&& J_{x(z)} = \mp \sqrt{\frac{(A-B)^2}{4}+C^2} + \frac{A+B}{2}, \label{eq9_3} \\
&& \Gamma_z = \left(h^z+2B\right)\cos(\alpha) + (\sqrt{2}h^x + 2C)\sin(\alpha) \label{eq10_3},\\
&& \Gamma_x = -\left(h^z+2B\right)\sin(\alpha) + (\sqrt{2}h^x + 2C)\cos(\alpha) \label{eq11_3}, \\
&& \label{eq:alphavstheta3} \alpha = 
\begin{cases}
    \frac{1}{2}\arctan\left(\frac{2C}{B-A}\right), & \text{if } 0 \leq \theta < \arcsin(\frac{2}{3}) \\
    \frac{\pi}{4},  & \text{if } \theta = \arcsin(\frac{2}{3}) \\
    \frac{1}{2}\arctan\left(\frac{2C}{B-A}\right) + \frac{\pi}{2},  & \text{if } \arcsin(\frac{2}{3}) 
 < \theta \leq \frac{\pi}{2}
\end{cases} \label{eq12_3}
\end{eqnarray}
for the parameters $A = 3\sin^2(\theta)-1$, $B = [3\cos^2(\theta)-1]/2$, and $C = 3\sin(2\theta)/(2\sqrt{2})$. These couplings are plotted vs $\theta$ in Fig.~\ref{fig:HamCouplings}.

The previous study of this model treated $h^z$ as a continuous parameter and did not consider a nonzero $h^x$. The study found that most of the interesting behavior involving the floating phases and both BKT and PT transitions occurred when $\Gamma_x=0$, as a nonzero $\Gamma_x$ breaks the $Z_2$ symmetry of the Hamiltonian. Accordingly, we will use both $h^z$ and $h^x$ to freely vary $\Gamma_z$ while ensuring that $\Gamma_x=0$ always. We do this by fixing $h^z=(\sqrt{2}h^x+2C)\cot(\alpha)-2B$, and tuning $\Gamma_z$ according to $\Gamma_z=-(\sqrt{2}h^x+2C)\csc(\alpha)-2B\cos(\alpha)$. With these changes to the model, we ensure that the Hamiltonian always has a $Z_2$ symmetry that corresponds to the parity of spins pointing up along the $z$-axis. This allows us to see a more complete picture of the Luttinger liquid behavior in the dipolar spin chain.

In this work, we are largely interested in quantum simulating the physics of Kitaev chains, so it is useful to consider Eq.~(\ref{eq:rotatedHam3}) in the spinless fermion basis. The mapping to the interacting Kitaev model is exact with OBC by rotating every other spin around its $z$-axis by $\pi$ and performing a Jordan-Wigner transformation to get 
\begin{eqnarray}
\label{eq:KitaevChain}
H &=& \sum_i -t(c^\dag_ic_{i+1} + c^\dag_{i+1}c_i) + \Delta(c_ic_{i+1} + c^\dag_{i+1}c^\dag_i)  \nonumber \\
 &+& U(2n_i-1)(2n_{i+1}-1) - \mu n_i,
\end{eqnarray}
where the hopping amplitude $t = J_x + J_y$, the pairing coefficient $\Delta = J_y - J_x$, the nearest-neighbor repulsion coefficient $U = J_z$, and the effective chemical potential $\mu = 2\Gamma_z$. Fig.~\ref{fig:HamCouplings} shows these parameters plotted vs $\theta$. For either representation, the Hamiltonian's parameters are all determined by only $\theta$ and $h^x$. This representation preserves the $Z_2$ symmetry, which now corresponds to the parity of the number of particles. There is no difference in the thermodynamic behavior of the spin representation vs. the fermionic representation with OBC or PBC; however, in the Kitaev representation, even-parity ground states with PBC have the interaction between sites $1$ and $N$ rotate like $t_{N,1} \rightarrow -t_{N,1}$ and $\Delta_{N,1} \rightarrow -\Delta_{N,1}$.

As this model describes a real system of dipolar-coupled spin centers implanted in a crystal, there are important considerations that must be addressed to realize this system experimentally. Within typical spin centers, hyperfine interactions can be large because of the close proximity between the electrons and the nuclei to which they are bound. The popular NV$^-$ spin centers can have hyperfine interactions with strengths of $\gtrsim 2$~MHz~\cite{DOHERTY20131,PhysRevB.94.155402,PhysRevB.100.075204}, which is larger than the $\sim 50$~kHz dipolar interaction strength between spin centers. However, other spin centers such as divacancies in SiC and SiV$^0$ centers in diamond do not present problematic hyperfine interactions because the majority of Silicon and Carbon isotopes are $^{28}$Si and $^{12}$C, respectively, which have zero nuclear spin. Furthermore, SiC and diamond can be grown to be isotopically pure to eliminate hyperfine couplings between the spin center electrons and the nuclei in the crystal~\cite{TERAJI2014231,doi:10.1126/sciadv.abm5912}.

Some challenges with implementing dipolar spin center chains cannot be immediately resolved. This includes implanting spin centers in the crystal to be between $5$ and $10$~nm, so that the dipolar interaction is maximized while the exchange interaction can be neglected. Moreover, the spin centers must have nearly even spacing, so that the disorder is not strong enough to break the physics that we present in this work. NV$^-$ centers have been implanted with a separation of $16 \pm 5$ nm~\cite{doi:10.1021/nl504441m}, which is nearing our requirements for this proposal, but is not there yet. Another challenge is that in temperature units, the dipolar interaction is $J/k_B\approx 2~\mu$K for spin centers separated by $10$~nm. The quantum simulator must operate at a temperature below this threshold, which is currently inaccessible in experimental setups. To resolve both implantation and temperature issues, we propose utilizing engineered interactions between spin centers that are stronger and longer-ranged. In particular, spin-spin interactions between spin centers that are mediated by bosonic modes have been proposed and include photonic~\cite{Bernien2013,doi:10.1126/science.1253512,Hensen2015}, polaritonic~\cite{PhysRevApplied.10.024011}, phononic ~\cite{PhysRevLett.117.015502,PhysRevLett.120.213603} and magnonic~\cite{trifunovic2013long,flebus2019entangling,zou2020tuning,candido2020predicted,neumanprl2020,fukami2021,peraca2023quantum,PhysRevB.106.L180406,main_hetenyi2022long,main_PhysRevB.105.245310}, mediated interactions. In Ref.~\cite{losey2024quantum}, the experimental considerations for a spin-$1/2$ dipolar-coupled spin center chain are described in more detail.

\section{Results}
\label{sec:results3}

To understand the phases and transitions in the dipolar spin center array, we used the \textsc{ITensor Julia Library}~\cite{10.21468/SciPostPhysCodeb.4} to perform finite-size density matrix renormalization group (DMRG) to calculate the ground state for various parameters. We use a truncation error of $10^{-10}$, while with each sweep, we gradually increase the maximum allowed dimension of the largest bond in the ground state matrix product state (MPS) to between $500$ and $750$. Following this, we use ground state MPSs and tensor network methods to calculate entanglement entropy and expectation values to determine the behavior of the spin chain. The von Neumann entanglement entropy $S_{\rm{vN}}$, which measures the degree of quantum entanglement between two partitions of a quantum many-body system, was used as a universal indicator of quantum phase transitions [see Fig.~\ref{fig:phaseDiagram}(a)]. The von Neumann entanglement entropy is defined as
\begin{eqnarray}
\label{eq15_3}
S_{\rm{vN}} = - \Tr[ \hat{\rho}_{\mathcal{A}} \ln(\hat{\rho}_{\mathcal{A}})],
\end{eqnarray} 
where the subscripts $\mathcal{A}$ and $\mathcal{B}$ refer to the two subsystems, and $\hat{\rho}_{\mathcal{A}} = \Tr_{\mathcal{B}} \hat{\rho}$ is the reduced density matrix. Furthermore, $\hat{\rho}$ is the density matrix of the entire system and is equal to $\ket{\Psi_0}\bra{\Psi_0}$ when the system is in the ground state $\ket{\Psi_0}$. Here, we always choose $\mathcal{A}$ to be half of the system when calculating $S_{\rm{vN}}$. Regions of the phase diagram with high entanglement entropy can indicate gapless phases or critical phase transitions. We confirm the criticality of these regions with the conformal field theory (CFT) prediction that the von Neumann entanglement entropy of a system diverges logarithmically with the system size as \cite{AffleckCritical1991,HOLZHEY1994443,VidalEntangle2003,PasqualeCalabrese_2004,Calabrese_2009}
\begin{eqnarray}
\label{eq:eescaling3}
    S_{\rm{vN}} = s_0 + \frac{c}{A}\ln (N),
\end{eqnarray}
where $s_0$ is a non-universal constant and $A = 3 \; (6)$ for PBC (OBC). The central charge $c$ helps determine the universality class and critical exponents associated with a phase transition, and can be extracted from the scaling of the entanglement entropy with the system size.

The phase diagram [see Fig.~\ref{fig:phaseDiagram}(a)] shows a low entanglement entropy region centered around $\Gamma_z = 0$. This region persists for $|\Gamma_z|<1.38$ when $\theta = \pi/2$, but as $\theta$ decreases, the value of $\Gamma_z$ that induces the transition out of the low entanglement entropy phase also decreases. This trend continues until $\theta = 0$, where the Hamiltonian is the Heisenberg point, a BKT point with SU(2) symmetry that is broken by any nonzero magnetic field or anisotropy. Fig.~\ref{fig:phaseDiagram}(b) shows the peak wave vector $k^z_p$ that describes the oscillations in correlations, such as $\langle S^z_iS^z_j\rangle$, and the spin density profile $\langle S^z_i\rangle$ with OBC. This peak wave vector is the value of $k$ that maximizes the Fourier transform of the spin density profile with the magnetization subtracted $\widetilde{S}^z_k = \sum_{j=1}^Ne^{-ikj}(\langle S^z_j\rangle-m^z)/N$. Fig.~\ref{fig:phaseDiagram}(c) shows the natural logarithm of the absolute value of $\widetilde{S}^z_{k_p}$. In the low entanglement entropy region described in this paragraph, $k^z_p/\pi=1$, which means that $\widetilde{S}^z_{k_p}$ is equal to the staggered magnetization order parameter, defined as $m^z_{stag} = [\sum_i (-1)^i \langle S^z_i\rangle]/N$, which shows that this low entanglement entropy phase is antiferromagnetic (AFM). This is consistent with studies of similar models \cite{losey2024quantum,chepiga2023eight}. The Heisenberg point is also antiferromagnetic, but differs because it has quasi-long-range order, unbroken SU(2) symmetry, and is described by a CFT with $c=1$.

There is another low-entanglement entropy region for large $\Gamma_z$. When $\theta=0$, this is a saturated ferromagnetic state where all spins point up along the $z$-axis~\cite{langari1998phase}. However, when $\theta \neq 0$, this is a disordered phase that has all spins almost pointing up along the $z$-axis. This is consistent with $k^z_p=0$ because an infinitely long wavelength on a finite-sized lattice is ferromagnetic. Moreover, $\widetilde{S}^z_{k_p}$ is very small in this region because the oscillations in the spin density profile have nearly zero amplitudes. The orange dashed line [see Fig.~\ref{fig:phaseDiagram}(a)] separating this low-entanglement region from the high-entanglement region is analytically known from previous studies of the interacting Kitaev model and in the spin basis is given by $\Gamma_z = \sqrt{(2J_z-Jx-Jy)^2 - (Jx-Jy)^2}$ \cite{katsura2015exact}.

\begin{figure*}[t!]
  \begin{center}
    \includegraphics[width=\textwidth]{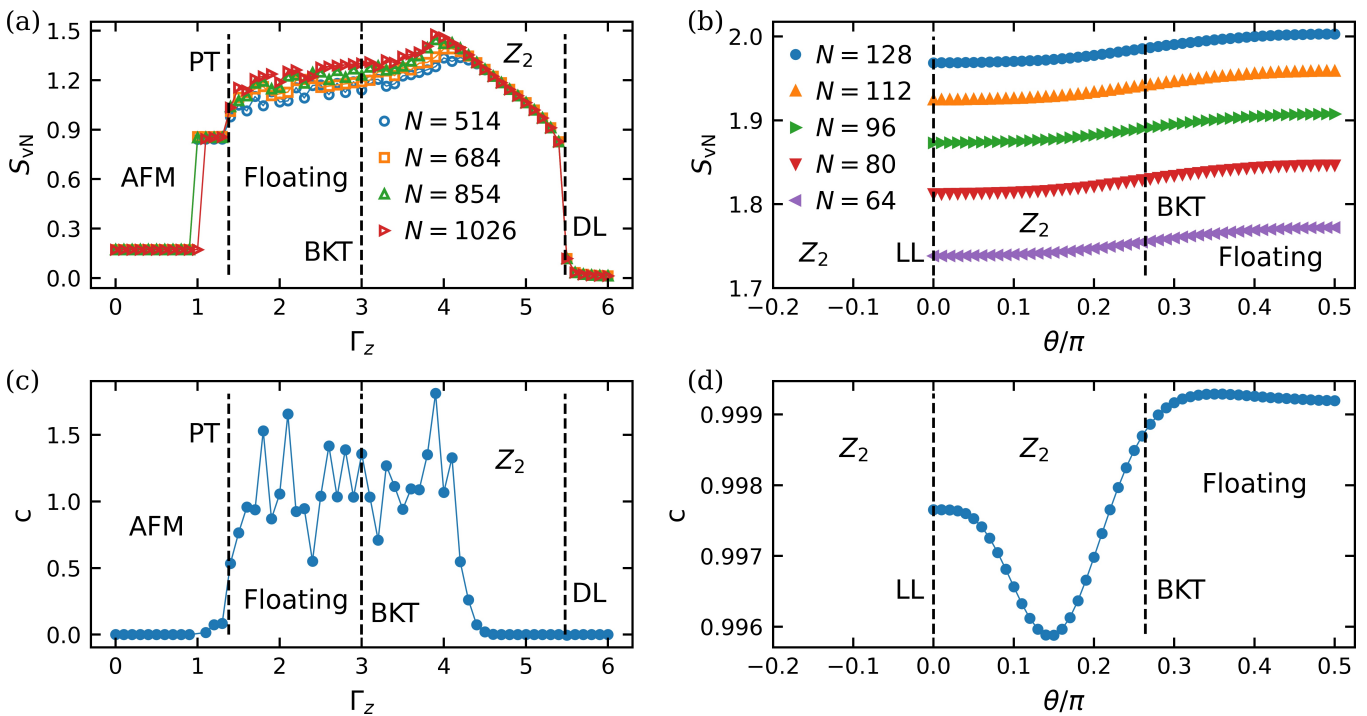}
    \caption{\label{fig:S1_c_cuts}Subplots (a) and (c) show entanglement entropy and central charge for a cut across the phase diagram where $\theta=\pi/2$. PT and BKT transitions are marked by black dashed lines at $\Gamma_z=1.38$ and $\Gamma_z=3.00$, respectively. Furthermore, the disorder line is marked by the black dashed line labeled DL at $\Gamma_z=5.477$. The phases are AFM to the left of the PT transition, floating between the PT and BKT transitions, $Z_2$ symmetry-breaking between the BKT transition and disorder line, and disordered to the right of the disorder line. For readability, subplot (c) doesn't show data points where $c>2$. Subplots (b) and (d) show entanglement entropy and central charge for a cut across the phase diagram where $m^z=1/16$ [see solid brown line in Fig.~\ref{fig:phaseDiagram}(a)]. This line begins in the pure Luttinger liquid phase at $\theta=0$ and crosses a BKT transition when $\theta = 0.26\pi$. Both of these points are marked by black dashed lines. The $Z_2$ symmetry-breaking phase is on either side of the point that belongs to the pure Luttinger liquid phase, and the floating phases exist on the right side of the BKT transition.}
  \end{center}
\end{figure*}

Between the two low entanglement entropy regions is a large region with high entanglement entropy. When $\theta \neq 0$, this region is composed of critical floating phases for smaller $\Gamma_z$ and a $Z_2$ symmetry-breaking phase for larger $\Gamma_z$. Both of these phases are incommensurate, as seen by the continuous changing of the peak wave vector $k^z_p$. The $Z_2$ symmetry that is broken corresponds to the parity of spins pointing up along the $z$-axis, in the spin picture, or the parity of the particle number in the fermionic picture. At $\theta = 0$, there is a gapless Luttinger liquid from the Heisenberg point at $\Gamma_z = 0$ to a PT point at $\Gamma_z = 4$. The $\theta = 0$ Luttinger liquid phase has a central charge $c=1$ and a microscopic $U(1)$ symmetry that conserves the $m^z$ magnetization in the spin picture and the number of particles in the Kitaev picture. This $U(1)$ symmetry arises from the fact that the coefficient for the pairing term $\Delta=0$ when $\theta=0$. While the floating phases are of the Luttinger Liquid universality class and also described by a conformal field theory with $c=1$, they have an emergent $U(1)$ symmetry rather than microscopic, since the coefficient for the pairing term is non-zero. The floating phases are separated from the AFM phase by a PT line with Luttinger liquid parameter $K=1/4$ and from the $Z_2$ breaking phase by a BKT line with $K=1/2$ [see Fig.~\ref{fig:phaseDiagram}(a)], since the scaling dimension of the pairing term is $1/K$, so the pairing term is irrelevant when $K<1/2$ \cite{chepiga2023eight}. This $Z_2$ breaking phase is closely related to a topologically non-trivial phase with gapless edge modes and winding number $w=1$ \cite{mondal2022detecting} that is of high-interest for quantum simulation and occurs when the interaction coefficient $U\rightarrow0$. In the remainder of this section, we locate the PT and BKT transition lines. To find the BKT transitions, we employ several methods to calculate the Luttinger liquid parameter $K$. Notably, this includes utilizing the energy spectrum at the PT transition to resolve an issue that arises when calculating $K$ from the crosscap overlap when the ground state is incommensurate.


The Tomonaga-Luttinger liquid (TLL) theory predicts most of the critical behaviors in 1-dimensional quantum many-body systems, including the floating phases. It says that by using bosonization to map our many-body model to a system of massless bosonic fields, we can completely describe the low-energy properties of the model with two parameters. These parameters are the effective velocity $u$ of the bosonic fields and the Luttinger liquid parameter $K$, which functions as a renormalized stiffness. The interactions between the bosonic fields are attractive for $K>1$ and repulsive for $K<1$, while the fields are non-interacting when $K=1$. In our model, the Luttinger liquid parameter lies within the range $1/4<K<1/2$ throughout the floating phases.

\begin{figure*}[t!]
  \begin{center}
    \includegraphics[width=\textwidth]{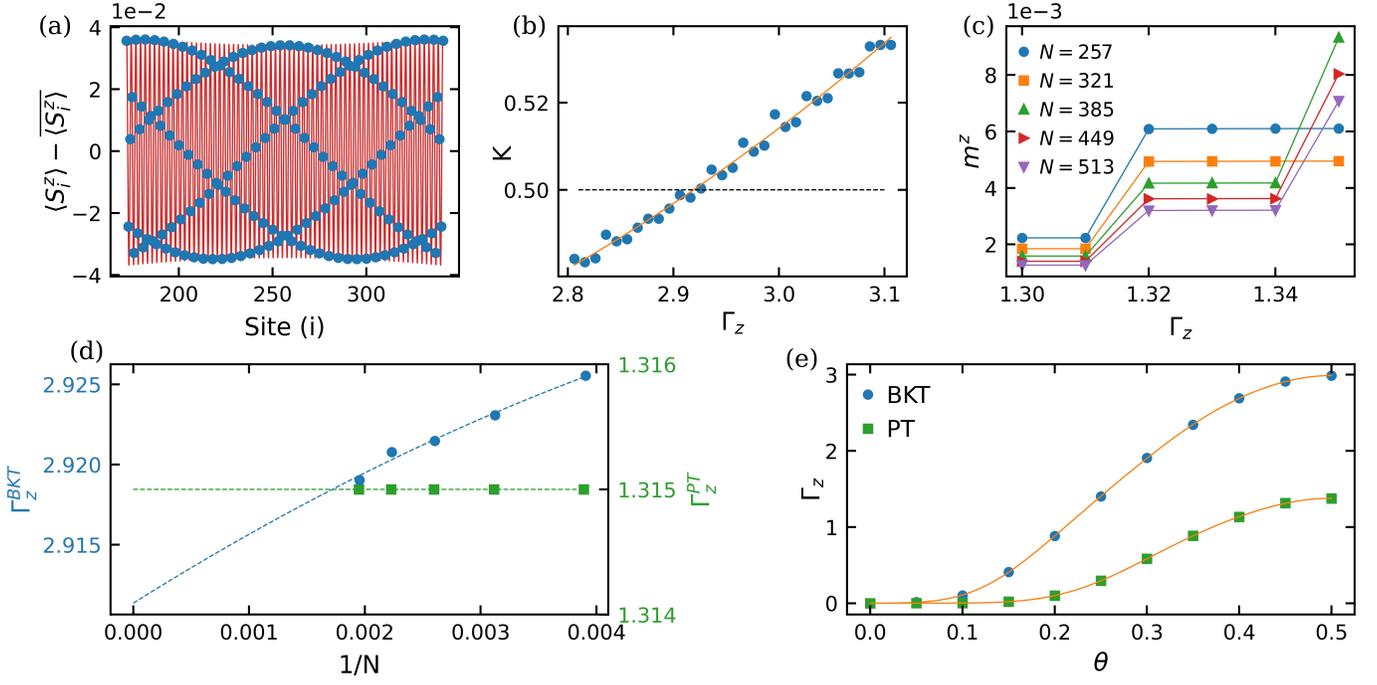}
    \caption{\label{fig:PT_BKT}Subplot (a) shows an example of a fit of the spin density profile used to extract the Luttinger liquid parameter $K$ from Friedel oscillations, when $N=512$, $\theta=0.45\pi$, and $\Gamma_z=2.906$. Subplot (b) shows a polynomial fit of $K$ vs $\Gamma_z$ when $N=512$ and $\theta=0.45\pi$. The BKT transition occurs when $K=1/2$. Subplot (c) shows the magnetization $m^z$ for various system sizes and for $\Gamma_z$ near the PT transition when $\theta=0.45\pi$. For these system sizes, all PT transitions fall within the range $1.31\leq \Gamma_z \leq 1.32$. Subplot (d) shows the procedure when $\theta=0.45\pi$ to locate the PT and BKT transitions in the thermodynamic limit. They are each fit to an inverse polynomial of $N$ and extrapolated to $1/N=0$. Subplot (e) shows the $11$ points where we located PT and BKT transitions in the thermodynamic limit. Each set of transitions is fit to a Fourier series to find an equation that describes the entire PT and BKT lines.}
  \end{center}
\end{figure*}

PT transitions occur from an incommensurate phase to a commensurate phase when the effective velocity $u$ of the bosonic fields goes to zero and the leading dispersion relation becomes quadratic with the momentum of low-energy excitations. If it is known between which phases PT transitions occur in a model, they are generally easy to locate with DMRG because there will be a sharp increase in the entanglement entropy between the gapped phase and the gapless Luttinger liquid phase. While that method performs well for our system with PBC or an odd number of sites, we find that magnetization is a better indicator of the PT transitions in our system when there are OBC and an even number of sites in the chain. This is because with OBC, the two edge sites are polarized in the direction of the magnetic field, since they each only interact with a single other site. As a result, as the magnetic field $\Gamma_z$ increases in the AFM phase with even N, just before reaching the PT transition, a soliton is created so that both edge sites can align with $\Gamma_z$. Interestingly, this creates an intermediate state between the AFM phase and floating phases that can be described as an AFM ground state with a domain wall, as it is gapped, yet incommensurate. It is not a real phase as it appears to disappear in the thermodynamic limit. However, this state can persist up to over $1,000$ sites, and the domain wall adds a constant $\ln(2)$ to the von Neumann entanglement entropy, so the sharp increase in entanglement entropy does not occur at the PT transition under these conditions. However, the floating phases always begin when increasing $\Gamma_z$ creates a pair of spinons, which together add $1$ to the total spin of the AFM ground state. Consequently, no matter the parity of $N$ or boundary conditions, we can precisely locate the PT transitions with magnetization. We only must consider that the AFM ground state near the PT transition has net spin $\sum_i S^z_i=1/2$ with odd $N$, $0$ with even $N$ and PBC, and $1$ with even $N$ and OBC [see Fig.~\ref{fig:PT_BKT}(c)].

BKT transitions occur between a gapless phase with incommensurate power-law decaying correlations and a disordered phase with exponentially decaying correlations when the Luttinger liquid parameter $K$ reaches a value where Luttinger liquid theory is unstable. In our model, we find a line of BKT transitions between the floating phases and the $Z_2$ symmetry-breaking phase.

Typical methods used to locate phase transitions are generally not adept at locating BKT transitions, due to the presence of marginal operators that often occur at BKT transitions. These marginal operators can create significant finite-size corrections to standard CFT equations. In our model, the BKT transition between the floating phases with emergent $U(1)$ symmetry and the $Z_2$ symmetry-breaking phase [see the purple dashed line in Fig.~\ref{fig:phaseDiagram}(a)] is where the pairing operator that explicitly breaks the $U(1)$ symmetry is marginal. For these phases, the central charge should change from $c=1$ in the LL universality class floating phases, to $c=0$ in the $Z_2$ symmetry-breaking phase, where $c=0$ means $S_\text{vN}$ saturates as $N$ increases [see Eq.(\ref{eq:eescaling3})]. However, one must be careful when analyzing the entanglement entropy for a finite-sized system in floating phases because the ground state will change discretely as the magnetization $m^z$ changes with system size. This means that to scale the system size to calculate quantities like central charge, one must consider specific system sizes and parameters for the Hamiltonian, so that the ratio of the $U(1)$ charge $n$ and $N$ remains constant for each ground state as $N$ changes, since $m^z=n/N$. Fig.~\ref{fig:S1_c_cuts}(a,c) shows that without keeping $m^z$ constant, the extracted value of $c$ fluctuates wildly because the entanglement entropy does not fit Eq.~(\ref{eq:eescaling3}) well. Furthermore, the entanglement entropy does saturate as expected in the $Z_2$ symmetry-breaking phase, but up to $1026$ sites, the magnetic field $\Gamma_z$ where the saturation begins does not closely align with the BKT transition. However, in Fig.~\ref{fig:S1_c_cuts}(b,d), we show that holding $m^z$ constant does make it possible to get extremely high quality fits for Eq.~(\ref{eq:eescaling3}) to precisely extract $c=1$ within the floating phases. Unfortunately, $c$ and $S\rm{_{vN}}$ remain ineffective for precisely locating the BKT transitions, as we also find $c=1$ in the $Z_2$ breaking phase for $N\approx 100$ with PBC. We do not use the scaling function for the energy gap at a BKT transition~\cite{dalmonte2015gap} to locate BKT transitions in this work, but believe that this method may be adept at locating the BKT transitions if special care is taken to keep the ratio of the magnetization fixed, like when extracting $c$.

As BKT transitions occur for a fixed value of the Luttinger liquid parameter $K$, extracting $K$ from the ground state wavefunction can be a potent way to locate BKT transitions. In a Luttinger liquid phase with OBC and polarized edge sites, Friedel oscillations in the expectation value of spin along the $z$-axis at each site $\langle S^z_i \rangle$ are described by 
\begin{eqnarray}
\label{eq:Friedel}
\langle S^z_i \rangle = \frac{A \cos\Bigl[k^z_p\Bigl(i-\frac{N+1}{2}\Bigr)\Bigr]} {\sin\Bigl(\frac{\pi i}{N+1}\Bigr)^K} + B, 
\end{eqnarray}
where $A$ is the amplitude of oscillations in $\langle S^z_i \rangle$ at the center of the chain, $B$ is the magnetization along the $z$-axis near the center of the chain, and $k^z_p$ is the floating phase wave vector, which is the wave vector for the incommensurate correlations~\cite{chepiga2022kosterlitz}. While care must be taken to ensure the fit is very high quality, this provides a relatively simple way to extract $K$. The largest source of error when extracting $K$ from Friedel oscillations oftentimes comes from the choice of how many edge sites to discard; however, the largest error in $K$ that we observe when removing between $1/3$ and $1/8$ of sites on each edge is $10^{-3}$. Fig.~\ref{fig:PT_BKT}(a) shows an example of a data fit with Eq.~(\ref{eq:Friedel}). Since the scaling dimension of the pairing term, which reduces the $U(1)$ symmetry to $Z_2$ is $1/K$, the BKT transition occurs when $K=1/2$. After locating the point where $K=1/2$ for different system sizes at fixed $\theta$ [see Fig.~\ref{fig:PT_BKT}(b)], we are able to locate BKT transitions by fitting these points to an inverse polynomial of $N$ and extrapolating to the thermodynamic limit. We perform the same extrapolation procedure for the PT transitions and provide an example of the extrapolations in Fig.~\ref{fig:PT_BKT}(d).

We identify the PT and BKT transitions in the thermodynamic limit for $11$ equally spaced values of $\theta$ in the range $0\leq \theta \leq\pi/2$, including the end points. We then fit the PT points and BKT points to a Fourier series in order to identify the entire PT and BKT lines in the phase diagram. The Fourier series provides a great fit here because the magnetic field that induces the PT and BKT transitions varies with the coupling constants in the Hamiltonian, which in turn vary with $\theta$. Moreover, the PT and BKT lines are even functions with respect to $\theta=\pi/2$, either even or odd functions with respect to $\theta=0$, and both intersect at the point $\Gamma_z=\theta=0$, which greatly reduces the number of viable terms to include in the Fourier series. With these constraints, we find high-quality data fits for $\Gamma_z^{PT}(\theta)=\sum_{j=0}^6d_jcos(2j\pi x)$ and $\Gamma_z^{BKT}(\theta)=\sum_{j=0}^6g_jsin([1+2j]\pi x)$, with coefficients given by the vectors $\mathbf{d}=[0.5019,\allowbreak -0.7021,\allowbreak 0.1950,\allowbreak 0.0210,\allowbreak -0.0106,\allowbreak -0.0076,\allowbreak 0.0023]$ and $\mathbf{g}=[2.4573,\allowbreak -0.6245,\allowbreak -0.1202,\allowbreak -0.0253,\allowbreak 0.0043,\allowbreak 0.0083,\allowbreak 0.0059]$. The Fourier series fit is shown in Fig.~\ref{fig:PT_BKT}(e).

\begin{figure}[t]
  \begin{center}
    \includegraphics[width=.48\textwidth]{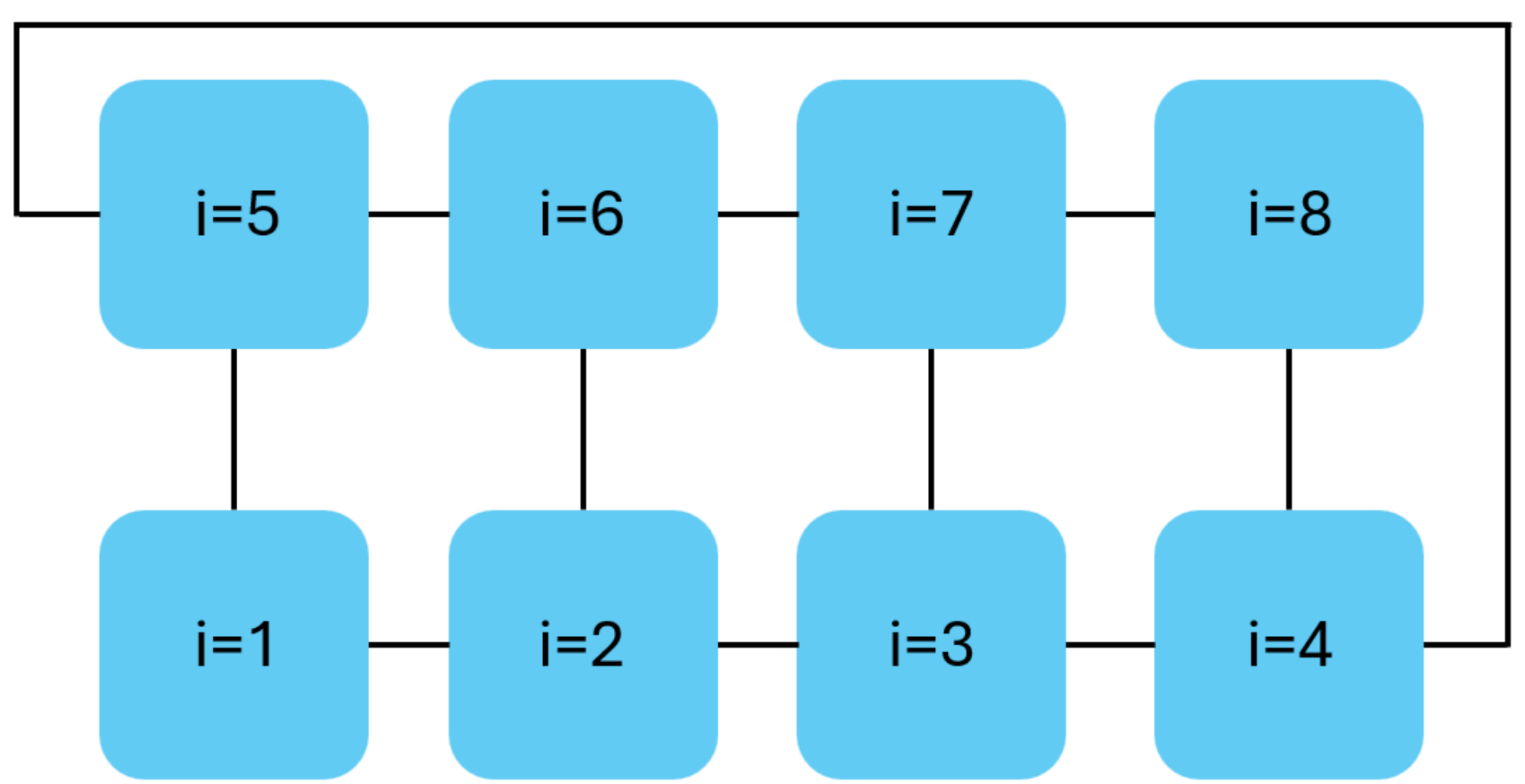}
    \caption{\label{fig:crosscapOverlap}Example of tensor contractions used to efficiently calculate the crosscap overlap with a spin-$1/2$ ground state matrix product state with $N=8$.}
  \end{center}
\end{figure}

We also investigate a novel method from conformal field theory for extracting the Luttinger liquid parameter $K$ by taking the overlap of the ground state wave function and a crosscap function~\cite{tan2025extracting}. For spin-$1/2$ chains with PBC, the crosscap function is given by 
\begin{eqnarray}
\label{eq:crosscap}
\hspace{-.6cm}|C_{\text{latt}}\rangle = \prod_{j=1}^{N/2} 
\left( |\uparrow\,\rangle_j |\uparrow\,\rangle_{j+N/2} 
+ |\downarrow\,\rangle_j |\downarrow\,\rangle_{j+N/2} \right).
\end{eqnarray}
When mod$(N,4)=0$, the following overlap gives 
\begin{eqnarray}
\label{eq:crosscapOverlap}
|\langle \psi_0|C_{\text{latt}}\rangle|^2=\frac{1}{\sqrt{K}},
\end{eqnarray}
where $|\psi_0\rangle$ is the ground state wave function. This overlap can be taken efficiently with matrix product states and thus pairs well with DMRG. Fig.~\ref{fig:crosscapOverlap} shows an example of how to calculate the crosscap overlap with tensor contractions for a spin-$1/2$ ground state matrix product state with $8$ sites. The procedure shown easily generalizes to larger system sizes, where the left half of the spin chain is always on the bottom and the right half of the spin chain is on the top. The dimension of the link between sites $N/2$ and $N/2+1$ can be large, so the memory cost of this calculation can be greatly reduced by separating this link into individual components when calculating the overlap.

\begin{figure*}[t]
  \begin{center}
    \includegraphics[width=\textwidth]{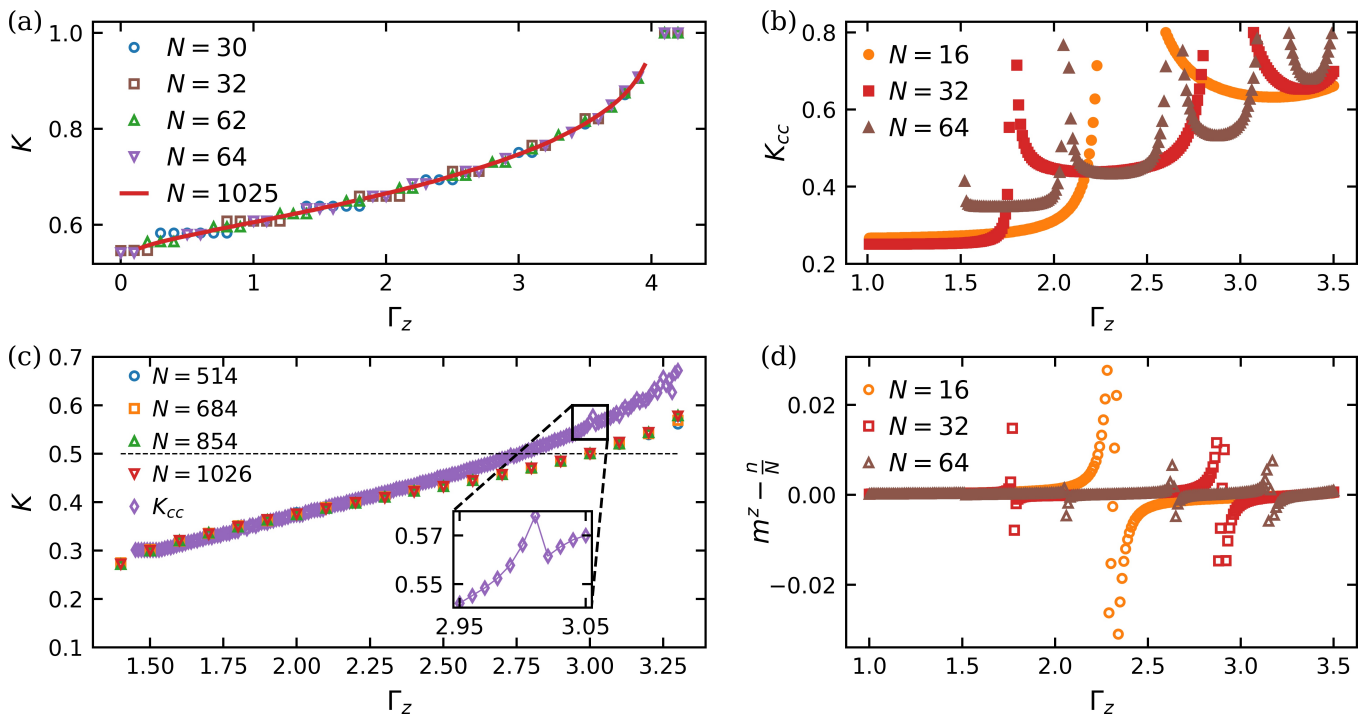}
    \caption{\label{fig:K}Subplot (a) shows the Luttinger liquid parameter $K$ extracted from the crosscap overlap for $N=30-64$ and from Friedel oscillations for $N=1025$, both when $\theta=0$, where the model has microscopic $U(1)$ symmetry. Subplot (b) shows $K$ extracted from the crosscap overlap $K_{cc}$ when $\theta=\pi/2$, where the model has no microscopic $U(1)$ symmetry and the ground state is incommensurate. Values of $K_{cc}$ larger than $0.8$ aren't shown for readability. Subplot (c) compares $K$ extracted from Friedel oscillations to $K_{cc}$ when $\theta=\pi/2$. For each $\Gamma_z$, $K_{cc}$ uses the energy spectrum at the PT transition to choose the system size that is expected to have the most nearly quantized $U(1)$ charge. The inset figure zooms in on a peak that occurs for $K_{cc}$ right where Friedel oscillations predict the BKT transition. Subplot (d) uses the magnetization to show how quantized the $U(1)$ charge is for each system size and $\Gamma_z$. $K_{cc}$ is unreliable when $|m^z-n/N|$ is large or corresponds to an odd parity ground state.}
  \end{center}
\end{figure*}

Since the crosscap state $C_{\text{latt}}$ only contains product states with even numbers of total spin, or equivalently particles, it is clear that the overlap $|\langle \psi_0|C_{\text{latt}}\rangle|=0$ when the ground state $|\psi_0\rangle$ has an odd total particle number $M$. This crosscap overlap is inversely proportional to the Luttinger liquid parameter $K$, so the method will give $K\rightarrow\infty$, and the method will have failed. However, when the ground state also has even $M$, the overlap gives $K$ according to Eq.~\ref{eq:crosscapOverlap}. This can further be understood by considering that $C_{\text{latt}}$ only has terms where every spin site $i$ is pointing in the same direction as its paired spin at site $i+N/2$. This equates the crosscap overlap to the tensor contraction shown in Fig.~\ref{fig:crosscapOverlap} and also to the reasoning that the overlap will be finite when the left half of the chain is "in phase" with the right half of the chain, and the overlap will be zero when the halves are "out of phase". Here, "in phase" essentially means that $\langle S^z_iS^z_{i+N/2}\rangle$ is a positive constant for all $i$. "Out of phase" then means that $\langle S^z_iS^z_{i+N/2}\rangle$ is a negative constant for all $i$. This reasoning agrees with the overlap being zero (non-zero) when $M$ is odd (even) because for the spin-1/2 XXZ model, where $M$ is exactly an integer for each ground state, the wavelength $\lambda_p$ for oscillations in the correlations $\langle S^z_iS^z_j\rangle$ is commensurate with $\lambda_p=N/\text{gcd}(N,N-M)$. When $N$ and $M$ are even, $\text{gcd}(N,N-M)$ is even, so each half of the chain is in phase with the other because each half of the chain will contain an integer number of wavelengths. However, when $N$ is even and $M$ is odd, $\text{gcd}(N,N-M)$ is always odd, so each half of the chain will be out of phase with the other because each half will contain a half-integer number of wavelengths. Fig.~\ref{fig:K}(a) shows excellent agreement when $\theta=0$ between $K$ extracted with Friedel oscillations when $N=1025$ and $K$ extracted with the crosscap overlap when $N$ and $M$ are even. The model has microscopic $U(1)$ symmetry when $\theta=0$, so the ground states are commensurate and confined to a single particle number sector.

Framing this overlap in terms of interference between oscillations from each half of the chain is useful reasoning and can be extended to explain troubles with the crosscap overlap when the oscillations are incommensurate. When the oscillations are incommensurate, each half of the chain won't be completely out of phase with each other and make $|\langle \psi_0|C_{\text{latt}}\rangle|=0$, but the magnitude of the overlap will be reduced and cause error by increasing the extracted value of $K$. In the thermodynamic limit, the oscillations will completely cancel each other and make $K\rightarrow\infty$. However, while this wave interference explanation fits so far, we will argue that there is a more subtle reason that is the true cause of why the overlap decreases when the wave function is a superposition of different particle number sectors, as in our floating phases.

Since all product states in $C_{\text{latt}}$ have different even numbers of particles and states with different particle numbers are orthogonal to each other, the overlap factorizes such that $\langle \psi_0|C_{\text{latt}}\rangle = \sum_{M\in \text{even}} \langle\psi_0^M|C^M_{\text{latt}}\rangle$, where the superscript $M$ refers to the projection of the state into the $M$ particle sector. In our model, the incommensurability in the floating phases is due to the fact that there is no $U(1)$ symmetry, so the ground states are superpositions of states in different particle number sectors with the same parities, since there is $Z_2$ parity symmetry. For each projected state $|\psi_0^M\rangle$, for an even parity ground state, $M$ is even, so each half of the chain is in phase, and $K$ can be extracted nicely for that projected state. However, when considering the overlap with the entire ground state, the contributions from different $|\psi_0^M\rangle$ can add or subtract from each other to change the value of $K$ extracted according to Eq.~\ref{eq:crosscapOverlap}. In Fig.~\ref{fig:K}(b), $K_{cc}$ begins to increase and eventually goes to infinity when contributions from different particle number sectors cancel each other out. Fig.~\ref{fig:K}(d) shows that where the ground state has a quantized particle number aligns with where $K_{cc}$ is minimized and expected to be accurate. Therefore, we find that we can calculate $K$ more precisely by choosing ground states that predominantly belong to a single particle number sector. Although they deviate slightly at the BKT transition, Fig.~\ref{fig:K}(c) shows reasonable agreement between the Friedel oscillation and crosscap overlap methods for extracting $K$ when we use ground states that are nearly confined to a single particle number sector. We show below how the energy spectrum at the PT transition can be utilized to select these ground states.

When $\Gamma_z$ for the ground state is maximally far from the closest values of $\Gamma_z$ that cause the ground state to be an equal superposition of two different particle number sectors, $K$ can be precisely extracted from the crosscap overlap. At this point, the ground state has maximal weight for a single particle number sector. Moreover, the magnetization $m^z$ closely resembles its thermodynamic value for that $\Gamma_z$, which reduces errors in $K$ due to discrete changes in the magnetization for finite system sizes, since $K$ is generally a function of magnetization rather than $\Gamma_z$. If we cannot identify these ideal values of $\Gamma_z$, Friedel oscillations will be much better suited than the crosscap overlap to extract $K$ within the floating phases. However, we show that with knowledge about which values of $N$ and $\Gamma_z$ result in floating phase ground states with a nearly quantized particle number or $U(1)$ charge $n=M-N/2$, it is feasible to use the crosscap overlap to extract $K$ within floating phases.

\begin{figure*}[t]
  \begin{center}
    \includegraphics[width=\textwidth]{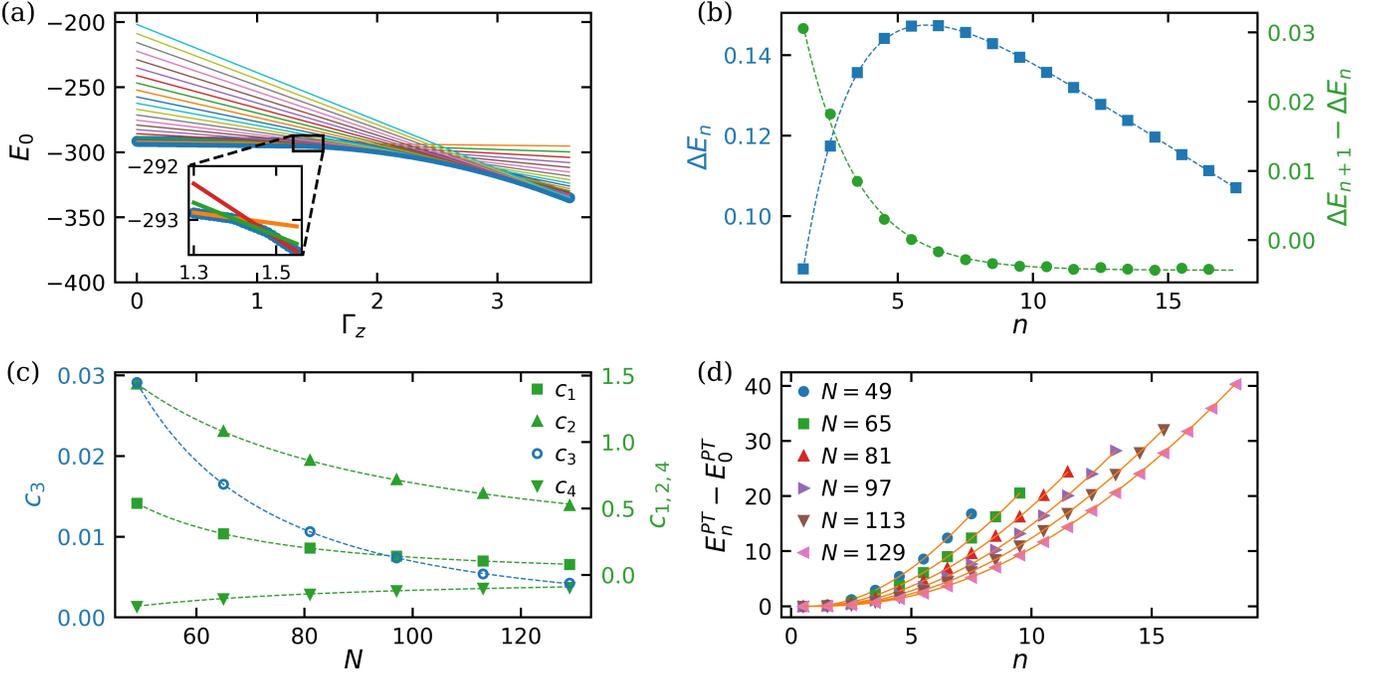}
    \caption{\label{fig:FPstepWidths}Subplot (a) shows the ground state energy as a function of magnetic field. There is a discrete set of ground states with different $U(1)$ charges $n$, as highlighted by the inset figure. The colorful lines are linear fits of the energy for each ground state. Subplot (b) shows the energy gap $\Delta E_n$ in the center of each discrete ground state, and how the energy gap changes between the centers of ground states. Subplot (c) shows the coefficients $c_i$ vs $N$. The dashed lines show the fits used to extract $a_i$ and $b_i$. Subplot (d) shows energies at the PT transition for states with different $U(1)$ charges and system sizes. The markers are data extracted from the linear fits in subplot (a) and the orange lines are plotted with Eq.~\ref{eq:EnPT} and the extracted $a_i$ and $b_i$.}
  \end{center}
\end{figure*}

We show that by extracting several parameters from calculated ground states, we can find the energy spectrum at the PT transition for ground states with any total particle number within the floating phases. Our model agrees with predicted scalings for large $N$, but also includes finite-size effects to give remarkably accurate results at small $N$. In our data, we observe that throughout the floating phases with $N$ held constant, the energy gap $\Delta E$ between the ground and first excited state changes for ground states with nearly quantized $U(1)$ charge $n$ as 
\begin{eqnarray}
\label{eq:DeltaEnGap}
\Delta E_{n+1} - \Delta E_n = c_1e^{-c_2n} - c_3,
\end{eqnarray}
where $c_1$, $c_2$, and $c_3$ are all positive and functions of the coupling constants and the system size. This means that for large $n$, the energy gap, which corresponds to the magnetic field needed to flip a spin from down to up in the $z$-basis, is proportional to the number of remaining down spins.

Summing all of these changes from $n$ to the maximum index for $\Delta E_{n+1} - \Delta E_n$ gives the energy gap for the floating phase state with nearly quantized $n$ as 
\begin{eqnarray}
\label{eq:EnGap}
\Delta E_n &=& c_4 - \sum_{m=n}^{N/2-1}(\Delta E_{m+1}-\Delta E_m) \\ &=& c_4 + \frac{c_1 \, \left( e^{-c_2 n} - e^{\tfrac{-c_2 \, N}{2}} \right)}{e^{-c_2}-1}
+ \, c_3 \, \left( \frac{N}{2} - n \right), \nonumber
\end{eqnarray}
where $c_4$ is negative, a function of the coupling constants and the system size, and corrects the error that comes from the ground states not having quantized $U(1)$ charge in the $Z_2$ breaking phase. Fig.~\ref{fig:FPstepWidths}(b) shows our data fit to Eq.~(\ref{eq:DeltaEnGap}) and Eq.~(\ref{eq:EnGap}). We find $\Delta E_n$ by finding where the slope of the ground state energy $E_0$ with respect to $\Gamma_z$ changes discretely [see Fig.~\ref{fig:FPstepWidths}(a)], since this slope is proportional to the $U(1)$ charge $n$ and the amount of magnetic field needed to transition between ground states with different $n$ is equal to the energy gap $\Delta E_n$. Fitting our data to extract $c_{1,2,3,4}$ for different $N$ and fitting each of them with $c_i=b_iN^{a_i}$ gives $a_1=-1.96$, $a_2=-1.03$, $a_3=-2.01$, $a_4=-1.00$, $b_1=1120$, $b_2=78.1$, $b_3=71.5$, and $b_4=-11.6$ [see Fig.~\ref{fig:FPstepWidths}(c)]. Rounding each $a_i$ to the nearest integer and expanding $\Delta E_n$ in powers of $1/N$ with fixed magnetization $m^z=n/N$ gives the dominant term at large $N$ as
\begin{eqnarray}
\label{eq:EnGapApprox}
\hspace{-.5cm} \Delta E_n \approx \frac{ 
b_{4} + \frac{b_1^{}}{b_2^{}} \left( e^{-b_{2}/2} - e^{-b_{2} \, m^z} \right)
+ b_{3} \left( \tfrac{1}{2} - m^z \right)} {N};
\end{eqnarray}
therefore, this result is consistent with the dynamical exponent $z=1$, which predicts $\Delta E\sim1/N$ for large $N$ within the floating phases.

To calculate the energy spectrum $E_n$ at the PT transition, we approximate the $U(1)$ as exactly quantized at the nearly quantized points, such that $E_n(\Gamma_z)=E_n^{PT}-2n\widetilde{\Gamma}_z$, where $\widetilde{\Gamma}_z=\Gamma_z-\Gamma_z^{PT}$, and $E_n^{PT}$ and $\Gamma_z^{PT}$ are the energy of the floating phase state with $n$ and the magnetic field both at the thermodynamic PT transition. Now knowing the energy gap where $n$ is nearly quantized gives the difference in energy between those states at the PT transition as $E_{n+1}^{\Gamma_z^{PT}} - E_{n}^{\Gamma_z^{PT}} = 2\widetilde{\Gamma}_z^{n,n+1}$, where $\widetilde{\Gamma}_z^{n,n+1}$ is again relative to $\Gamma_z^{PT}$, but also where the $n$ and $n+1$ states are degenerate. Therefore, $E_n^{PT} - E_{n_0-1}^{\Gamma_z^{PT}} = 2\sum_{m=n_0}^{n}\widetilde{\Gamma}_z^{m-1,m} = 2\sum_{m=0}^{n-n_0}(n-m-n_0+1)\Delta E_{m+n_0-1}$, where $\Delta E_{n_0-1} \equiv \widetilde{\Gamma}_z^{n_0-1,n_0} \sim 1/N^2$ at large $N$ because the critical exponent $\nu=1/2$ at PT transitions, while all other $\Delta E_n$ are given by Eq.~\ref{eq:EnGap}. With odd $N$, $n_0=3/2$, and with even $N$, $n_0=2$ with OBC or $n_0=1$ with PBC, as $n_0$ is just the smallest $U(1)$ charge where the ground state is within the floating phases. Performing the summation gives the energy spectrum at the PT point as 
\begin{widetext}
\begin{eqnarray}
\label{eq:EnPT}
E_n^{PT} - E_{n_0-1}^{PT} &=& 2(n-n_0+1)\Delta E_{n_0-1} + 
\frac{2c_1e^{c_2(2-n-n_0)}}{\big(e^{c_2}-1\big)^3}
\Big[
(e^{c_2}-1)(n_0-n)+1)e^{c_2n}-e^{c_2n_0}
\Big]
\\[4pt]
&+&\frac{c_1(n-n_0)(1+n-n_0)e^{c_2(1-\tfrac{N}{2})}}{e^{c_2}-1}
+(n-n_0)(1+n-n_0)\left(
c_4-\frac{c_3}{6}(2n+4n_0-3N-2)
\right). \nonumber
\end{eqnarray}
\end{widetext}

In Fig.~\ref{fig:FPstepWidths}(d), we show that using the values of $a_{1,2,3,4}$ and $b_{1,2,3,4}$ extracted in Fig.~\ref{fig:FPstepWidths}(c), we can precisely predict $E_n^{PT}-E_0^{PT}$ for various $n$ and $N$, after fitting for a single value of the proportionality constant for $\Delta E_{n_0-1} \sim 1/N^2$. Now expanding $E_n(N)$ in powers of $1/N$ for fixed $m^z$ and keeping the dominant term for large $N$, then taking $m^z=n/N$ to be small, we get
\begin{eqnarray}
\label{eq:EnPTApprox}
E_n^{PT}(N) \approx \left( \frac{b_{3}}{2} + b_{4} + \frac{b_{1}\left(e^{-b_{2}/2}-1\right)}{b_{2}} \right) \frac{n^2}{N},
\end{eqnarray}
which is consistent with the scaling given by the expectation $E_n^{PT}(N) \sim n^2/N$ when $n/N$ is small and $N$ is large~\cite{francesco2012conformal}. We find that the extra corrections when $N$ is not large are necessary to precisely calculate $E_n^{PT}(N)$ for any system size and $n$ that has a ground state within the floating phases. Now knowing the energy spectrum at the PT transition, we can determine for any system size, where the ground state in the floating phases transitions to a state with a different $U(1)$ charge with $\widetilde{\Gamma}_z^{n,n+1} = (E_{n+1}^{PT}-E_n^{PT})/2$. Furthermore, for a given $N$ and $n$, we can locate the $\Gamma_z$ where the ground state has $n$ is nearly quantized by taking the average of $\widetilde{\Gamma}_z^{n,n+1}$ and $\widetilde{\Gamma}_z^{n-1,n}$. While our expression for $E_n^{PT}(N)$ is not expressed explicitly in terms of the coupling constants, since the floating phases have emergent $U(1)$ symmetry, in practice $E_n^{PT}(N)$ can be numerically calculated by simply removing the Hamiltonian term that breaks the microscopic $U(1)$ symmetry and calculating the ground state at the PT transition with different particle numbers. Then $c_{1,2,3,4}$ and $\Delta E_0$ can be extracted to find $E_n^{PT}(N)$ for any $N$. Finding when the ground state has a nearly quantized $U(1)$ charge has many uses for floating phases, such as extracting central charge, potentially locating BKT transitions with an energy gap scaling function, or, as we show now, calculating the Luttinger liquid parameter $K$ with the crosscap overlap method.

Using Eq.~(\ref{eq:EnPT}), for every $0.01$ $\Gamma_z$, we calculate the system size in the range $32 \leq N \leq 128$ that will have a ground state with nearly quantized $n$ at that $\Gamma_z$. In Fig.~\ref{fig:K}(c), we now get a smooth line for $K$ throughout the floating phases. The line only becomes more jagged after leaving the floating phases near $\Gamma_z \approx 3$. The crosscap method appears to slightly overestimate $K$ in the floating phases, which may be due to error in how confined the ground state is to a single particle number sector, or due to corrections for the crosscap method, which has been shown to sometimes overestimate $K$ near BKT transitions~\cite{tan2025extracting}. Interestingly, we observe a slight peak in $K$ at the BKT point [see inset in Fig.~\ref{fig:K}(c)], which may be attributed to the method's breakdown outside the floating phases. Alternatively, this peak could also be due to extra corrections at the BKT transition, which could be utilized to locate BKT transitions.

\section{Conclusions}
\label{sec:conclusion3}

We have shown that a 1D chain of dipolar-coupled $S=1$ spin centers with anisotropy and no hyperfine interactions can quantum simulate the interacting Kitaev chain. Controlling the orientation of the spin center chain within the crystal and tuning external magnetic fields makes it possible to quantum simulate critical floating phases, a $Z_2$ symmetry-breaking phase, and both PT and BKT transitions. Furthermore, we show that entanglement entropy and central charge, which are considered universal indicators for many quantum phase transitions, are not reliable indicators for the BKT transitions between the floating and $Z_2$ breaking phases. However, the Luttinger liquid parameter is a reliable indicator for this phase transition, and we explore several methods for extracting it from ground states.

This proposal is for one of the first spin center-based quantum simulators and couples spin centers via the magnetic dipole-dipole interaction. Employing stronger and longer-ranged engineered interactions between spin centers can alleviate several remaining experimental challenges and make these devices feasible to build and operate. Moreover, this will unlock the possibilities to engineer 2D and 3D arrays of spin-$S$ spin centers with simple or complex geometries, to simulate a plethora of rich quantum behaviors, including spin dynamics and topologically protected edge modes.

\section{Acknowledgments}

Computations were performed using the computer clusters and data storage resources of the HPCC, which were funded by grants from NSF (MRI-2215705, MRI-1429826) and NIH (1S10OD016290-01A1). J.Z. is supported by NSFC under Grants No. 12304172 and No. 12347101, Chongqing Natural Science Foundation under Grant No. CSTB2023NSCQ-MSX0048 and No. CSTB2024YCJH-KYXM0064. This work was supported in part by the U.S. Department of Education through the Graduate Assistance in Areas of National Need (GAANN) program.

\bibliography{NVKitaevchainV2.bib}

\end{document}